\documentclass{mem}
\usepackage{natbib}\usepackage{txfonts}\usepackage{balance}
\usepackage{graphicx}
\usepackage[a4paper]{hyperref}
\idline{79}{1}
\begin{document}
\def\teff{$T\rm_{eff }$}
\def\kms{$\mathrm {km s}^{-1}$}

\title{Structure and Kinematics of the Broad-Line Region and Torus of
Active Galactic Nuclei}


\author{C.~Martin Gaskell\inst{1}, Ren\'e W. Goosmann\inst{2}, and
Elizabeth S.~Klimek\inst{3}}

\offprints{M.~Gaskell}

\institute{ Department of Astronomy, University of Texas, Austin, TX
78712-0259. \email{gaskell@astro.as.utexas.edu}
 \and Astronomical Institute of the Academy of Sciences, Bocni II
1401, 14131 Prague, Czech Republic \and Department of Astronomy, New
Mexico State University, Las Cruces, NM 88003-0001 }

\authorrunning{Gaskell, Goosmann, \& Klimek}

\titlerunning{BLR Structure and Kinematics}

\abstract{Energetics considerations imply that the broad-line region
(BLR) has a high covering factor.  The absence of absorption from
the BLR means that the BLR has to have a flattened distribution and
be seen through a polar hole.  The BLR is the inward extension of
the torus and they have similar geometries and covering factors.
Reconciling velocity-resolved reverberation mapping,
spectropolarimetry, and the increasing blueshifting of BLR lines
with decreasing distance from the centre, implies that the BLR has a
significant inflow component. This inflow provides the mass inflow
rate needed to power the AGN. We suggest that the mechanism
producing the outward transport of angular momentum necessary for
the net inflow of the BLR is the magneto-rotational instability, and
that the BLR and outer accretion disc are one and the same.
\keywords{accretion: accretion disk --- black hole physics ---
galaxies:active --- galaxies:quasars:emission lines
--- scattering} } \maketitle{}

\section{Introduction}

Even though active galactic nuclei (AGNs) are too far away and their
inner regions too small to resolve, many cartoons of suggested AGN
structures have nevertheless been published over the last four
decades or so. A {\it Google Images} internet search found cartoons
suggesting a very wide variety of possible structures.\footnote{The
only point they were unanimous on was that the black hole was always
in the centre!} This lack of a consensus is significant because it
can be argued that if we don't know what something {\it looks} like,
then we don't really understand it.

    Our best probe of the inner workings of AGNs is provided by the broad-line
region (BLR), but, unfortunately, one of the areas in which cartoons
of AGNs differ most is in the location and structure of the BLR.  It
is most commonly thought of as having a spherical or quasi-spherical
distribution, but it is also depicted as being between the accretion
disc and the torus, just above the accretion disc, or in cones
around the radio jets. Closely related to the question of where the
BLR gas is located is the question of how the BLR gas is {\em
moving}. Knowing the answers to these two fundamental questions is
crucial for understanding what the BLR is and for using it to
determine black hole masses. For example, starting with
\citet{dibai81}, for a given  observed BLR line width, $v$, and
optical luminosity, $L_{opt}$, the quantity $v^2L_{opt}^{1/2}$ has
been used to estimate black holes masses, but for such mass
estimates to be reliable we need to know both the distribution of
the BLR gas, and what is controlling the gas motions.\footnote{One
cannot, for example, get the mass of the sun by measuring the speed
of the solar wind!}

    In this paper we briefly discuss our recent results on the structure
and kinematics of the BLR. Further details can be found in
\citet{gaskell+08b}(= GKN) and \citet{gaskell_goosmann08}.

\section{The structure of the broad-line region}

\subsection{The BLR is flattened}

    If the spectral energy distribution of an ionizing continuum source
and the covering factor of the surrounding gas are known, the
equivalent widths of emission lines can be calculated from the
conservation of energy \citep{stoy33}. \citet{macalpine81} pointed
out that the equivalent width of the He~II line in AGNs implied a
very BLR high covering factor. In fact, based on the extrapolation
of reported UV continua shapes of the time, the covering factor,
$\Omega /4\pi$, needed seemed to be much greater than 100\%.  This
problem has been referred to as ``the energy-budget problem'' (see
\citealt{macalpine03} for a recent review).  Even a covering factor
of 100\% is too high because unambiguous Lyman continuum absorption
edges at $\lambda$912 due to the BLR are {\it never} seen (see
\citealt{koratkar_blaes99} for a review). If, as is widely assumed,
the BLR is spherically symmetric, the limits on Lyman limit
absorption put the BLR covering factor at only a few percent at
most, which is far too low to explain the observed equivalent widths
of of BLR lines.

NGC~5548 probably has the best studied BLR and continuum of any AGN
because it has been the target of intensive multi-wavelength
monitoring campaigns by the {\it International AGN Watch} ({\it
IAW}; e.g., \citealt{clavel+91}, \citealt{peterson+91},
\citealt{peterson+92}, \citealt{korista+95}) and other groups over
the past two decades. \citet{gaskell+08b} therefore chose NGC~5548
for a detailed analysis of BLR energetics.

\citet{macalpine81} suggested that reddening could be a solution to
the energy-budget problem. However, \citet{gaskell+04},
\citet{czerny+04}, and \citet{gaskell_benker08} have found that most
AGNs have relatively flat reddening curves, and the reddening curve
\citet{gaskell+08b} find for NGC~5548 is consistent with the mean
AGN reddening curve of \citet{gaskell_benker08}.  Dereddening by
this curve lowers the covering factor a little, but still leaves it
around $\sim 40$\% for lines of a wide variety of ionization.

Since variability leaves no doubt that BLR emission is driven by
photoionization, and the spectral energy distribution is fairly well
constrained, GKN argue that the combination of the high covering
factor needed to explain the BLR line strength, and the lack of
Lyman limit BLR absorption require that the BLR {\it cannot} be
spherically symmetric.  Instead {\em we have to always be viewing
the BLR through a hole.}  This has important consequences.

\subsection{The BLR is self-shielding}

The current standard model of the BLR has random clouds each with a
clear line of sight to the central ionizing source.  A typical
optically-thick cloud will produce high-ionization lines from its
front side and low-ionization lines from its back side. Each cloud
will thus produce lines through the full range of ionization,
although the ratios will change somewhat with the distance from the
ionizing source.  A random mix of such clouds (the so-called ``LOC''
model of \citealt{baldwin+95}) naturally reproduces the integrated
BLR spectrum.

However, because of the high covering factor ($\sim 40$\%) and the
requirement of a flattened BLR, {\em there is 100\% covering in the
equatorial plane}.  Thus the more distant clouds do {\em not} have a
clear line of sight to the central ionizing source. Instead, clouds
will self shield each other so that the radiation reaching the outer
BLR clouds will have been absorbed by going through the inner BLR.
This means that the more distant clouds behave like the back sides
of exposed clouds in the standard model, and the range in ionization
seen in a single cloud is now spread out across the whole BLR.  Such
a self-shielding BLR was explored by \citet{macalpine72}.

It has long been recognized that higher-ionization lines arise
closer in to the centre of an AGN on average
\citep{gaskell_sparke86,clavel+91}. The GKN model predicts a wider
range of effective radii than the standard model because in the GKN
model different clouds emit different lines, while an exposed cloud
LOC-type model predicts a narrower range of radii because every
cloud emits every line.  Fig. 1 shows the predictions of the two
models versus measured lags from the {\it IAW} monitoring of
NGC~5548. As can be seen, the self-shielding GKN model is a
significantly better fit.

\begin{figure}[]
\resizebox{\hsize}{!}{\includegraphics[height=80mm]{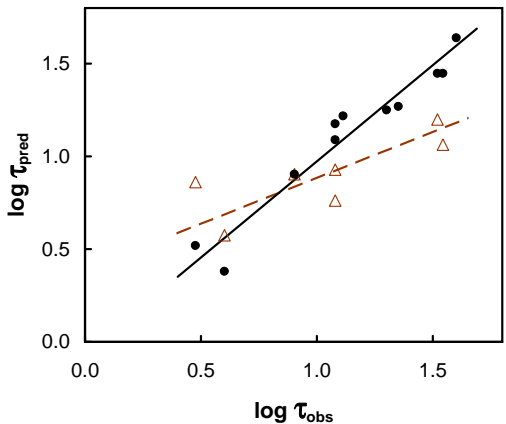}}
\caption{ \footnotesize Observed BLR reverberation mapping lags for
different ions in NGC~5548 versus the predicted lags from the
exposed cloud LOC model (brown triangles), and the self-shielding
model of \citet{gaskell+08b}.  The lines are OLS-bisector fits.}
\label{small_cartoon}
\end{figure}

\subsection{The BLR is the inner extension of the torus}

The BLR shields not only its own outer regions but also the torus,
and this needs to be allowed for when considering torus energetics.
Shielding means that (a) the torus has to have a larger covering
factor in order to reprocess less energy, and (b) the dust can
survive closer to the black hole because it receives less energy.
After allowance for the shielding, the torus opening angle
\citet{gaskell+08b} find for NGC~5548 is similar to the torus
opening angles estimated from the fraction of type-1/type-2 AGNs of
comparable luminosity

We believe that the BLR and torus covering factors are similar for
other AGNs in general. If the BLR has a larger covering factor, then
as we approach edge-on viewing, the BLR would start to block the
continuum before the torus did, yet, as noted above, we never see
BLR absorption. When we do see line absorption (in
broad-absorption-line quasars) it is always blueshifted and is
therefore part of a high-velocity outflow.  For thin tori seen near
edge-on we could also see strong BLR lines without seeing the AGN
continuum directly, and this also is never observed. Conversely, the
torus covering factor cannot be greater than the BLR covering factor
close in because the dust cannot survive there in the direct
radiation field of the AGN.

\citet{netzer_laor93} proposed that the outer boundary of the BLR is
set by dust formation and this is confirmed by IR reverberation
mapping \citep{suganuma+06}.  The BLR thus starts where the torus
stops, and, for the reasons just explained, they will have similar
covering factors at this point. We therefore believe that {\em the
torus and BLR are part of the same thing}.  For additional
discussion of this, see the contributions by Moshe Elitzur and Hagai
Netzer in these proceedings.

\section{The kinematics of the broad-line region}

Chaotic virialized motions, Keplerian orbits, radiatively-driven
outflows, infall, and various combinations of these possibilities,
have all been extensively considered for BLR motions. The dominant
early view was radial outflow. Radiation pressure accelerated lines
could produce the observed profiles \citep{blumenthal_mathews75},
blueshifted intrinsic absorption lines prove that at least some gas
is outflowing, and the discovery of the blueshifting of the
high-ionization BLR lines \citep{gaskell82} has been widely taken as
evidence of the outflow of the high-ionization BLR gas.

Unfortunately, velocity-resolved reverberation mapping
\citep{gaskell88} gave the conflicting result that the BLR was {\em
not} outflowing but gravitationally bound instead.  Because this
makes the current AGN black hole mass estimating industry possible
this is now a very popular idea.  However, the conflict with the
evidence for outflow, especially the blueshifting of the
high-ionization lines, has remained a major problem. The most widely
adopted solution has been to assume that the BLR has {\em two}
components (see \citealt{gaskell00} for a review) with different
kinematics: a gravitationally-bound, low-ionization BLR, and a
radially-outflowing, high-ionization BLR.  These two components have
often been associated with ``disk'' and ``wind'' components.
Unfortunately, as \citet{gaskell_goosmann08} point out, the first
velocity-resolved reverberation mapping showing no outflow was of
the {\em high}-ionization C\,IV line.

\citet{gaskell_goosmann08} show that the blueshifting can be
produced from electron scattering from an inflowing medium. There
are a couple of independent lines of evidence pointing to a net
inflowing component of the BLR velocity. Firstly, velocity-resolved
reverberation mapping studies have {\em repeatedly} shown that there
is an {\em infalling} component to the velocity
\citep{koratkar_gaskell89,crenshaw_blackwell90,koratkar_gaskell91a,koratkar_gaskell91b,korista+95,
done_krolik96,welsh+07}.  While this might be a marginal result for
just one line in one AGN, the overall significance for the entire
sample considered in the studies is substantial.  The second line of
evidence for infall comes from high-resolution spectropolarimetry:
the systematic change in polarization as a function of velocity
across the Balmer lines requires a net inflow of a scattering region
exterior to the Balmer lines \citep{smith+04}. On the basis of these
reverberation mapping and spectropolarimetric results,
\citet{gaskell_goosmann08} adopt net infall velocities of $\sim
1000$ km s$^{-1}$ (about a quarter of the FWHM of a typical BLR
line), and show, using the Monte Carlo radiative transfer code {\it
STOKES} \citep{goosmann_gaskell07}, that electron scattering off
infalling material produces the blueshifting of the high-ionization
lines.  Polarization reverberation mapping \citep{gaskell+08a} shows
that there is significant scattering at the radius of the BLR. As we
illustrate in Fig. 2, a range of optical depths and geometries for
the infalling scatterers readily reproduces the observed
high-ionization line profile.  The model also reproduces the
dependence of the blueshifting on the ionization of the line.

\begin{figure}[]
\resizebox{\hsize}{!}{\includegraphics[height=80mm]{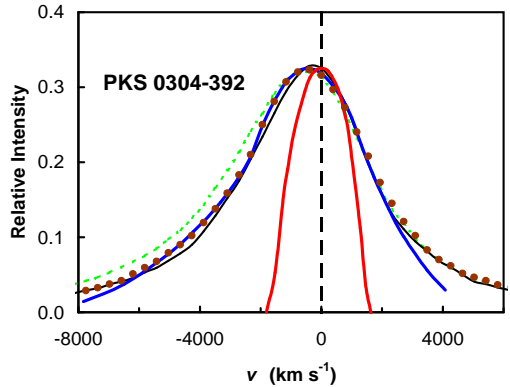}}
\caption{ \footnotesize The profiles of O\,I $\lambda$1305 (narrow
symmetric profile shown in red) and C\,IV $\lambda$1549 (thick blue
line) for the quasar PKS~0304-392.  The thin black line is the
blueshifted profile produced by a spherical distribution of scatters
with $\tau$ = 0.5, and the dashed green line is the profile produced
by the same distribution with $\tau$ = 1.  The brown dots are the
profile produced by an infalling cylindrical distribution of
scatterers with $\tau$ = 20 and a half opening angle of 45 deg.
PKS~0304-392 observations from \citet{wilkes84}}
\label{blueshift_example}
\end{figure}

\section{Discussion}

\subsection{The overall picture}

Our overall picture of an AGN is summarized in our own AGN cartoon
shown in Fig. 3. Material from the torus has a net inflow. As it
spirals in, the dust sublimates and we have a dust-free BLR. Because
the BLR and torus are physically thick there must be a vertical
``turbulent'' component of motion, $v_{turb}$ (see
\citealt{osterbrock78}).  At any given radius the relationship
between the various velocity components is

\begin{equation}
v_{Kep} > v_{turb} > v_{inflow} \approx 0.25 v_{Kep}
\end{equation}

\noindent where $v_{Kep}$ is the Keplerian orbital velocity.

To complete the picture, there is a higher velocity, low-density
outflow in the polar cones which is indicated by the dotted arrows
in Fig. 3 and is discussed elsewhere in these proceedings.  If we
animate our cartoon then in a co-rotating reference frame we would
see turbulent motions of the BLR with a slower net infall (the white
arrows in Fig. 3).

\begin{figure}[]
\resizebox{\hsize}{!}{\includegraphics[height=85mm]{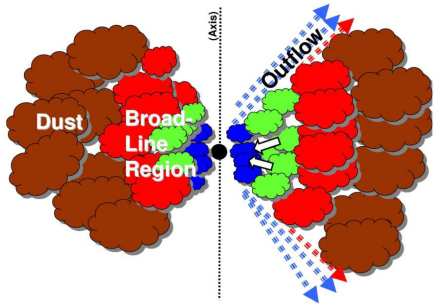}}
\caption{ \footnotesize A cartoon showing a cross section of an AGN
through the rotation axis.} \label{small_cartoon}
\end{figure}

\subsection{The mass inflow rate}

Since we have estimates of the inflow velocity, covering factor, gas
density, and BLR radius we can also estimate the mass inflow rate.
This is indeed similar to the accretion rate needed to power the
AGN.  We therefore assert that the BLR {\em is} the matter being
accreted onto the black hole.

\subsection{High accretion rate AGNs}

Our {\it STOKES} modelling of scattering off infalling regions shows
that the blueshifting increases with increasing column density (see
Fig. 2). It also depends, of course, on the inflow velocity. The
product of the column density and the inflow velocity gives the {\em
mass accretion rate}.  We therefore predict that high-accretion-rate
AGNs (narrow-line Seyfert 1s; NLS1s) will show higher blueshiftings.
Such large blueshifts, formerly interpreted as strong winds, have
indeed been found in NLS1s
\citep{sulentic+00,xu+03,leighly_moore04}.

\subsection{The BLR is the accretion disc}

The net inflow implies that

(a) there is a mechanism (a ``viscosity'') to transport angular
momentum outwards and take away energy.

(b) energy is being generated as the BLR falls in.

The need for enough viscosity to get the observed inflow is the same
as for classical accretion discs.  It is now recognized that this
viscosity comes from the magneto-rotational instability (MRI;
\citealt{balbus_hawley91}).  We propose that the MRI is also driving
the inflow of the BLR.

Our deduced motions of the BLR shown in Fig. 3 and described in
section 4.1 are qualitatively the same as the motions shown by MHD
simulations of accretion discs (see, for example, the movies
described by \citealt{hawley_krolik01}). Furthermore,
\citet{gaskell08} has pointed out that the size of the outer regions
of a typical accretion disk is comparable to the size of the BLR.
Given these similarities, we propose that the BLR {\em is} the outer
accretion disc.

\subsection{The narrow line region}

As is discussed elsewhere in these proceedings, there is no doubt
that {\em some} of the narrow-line region (NLR) is associated with
the outflow in AGNs. However, the NLR also shows increasing
blueshiftings with ionization just like the BLR, so Occam's razor
suggests that much of the NLR is also inflowing.

\section{Summary}

Putting the results discussed above together we end up with the
following picture of the AGN phenomenon:

\begin{itemize}

\item The BLR is the outer accretion disc.

\item The BLR is the inner part of the torus.

\item The BLR material flows in from the torus, through the BLR, and
onto the black hole.

\item The BLR motions are dominated by gravity, therefore $v^2
L^{1/2}$ can safely be used to estimate black hole masses.

\item Magnetic fields also play a role in BLR kinematics, but
radiation pressure is relatively unimportant.

\end{itemize}

\begin{acknowledgements}

Big thanks are due to Manolis Angelakis, Andrei Lobanov, and the
other organizers for their efforts in putting the meeting together.
It is a pleasure to acknowledge many useful discussion with Ski
Antonucci. This research has been supported in part by the US
National Science Foundation through grant AST 03-07912, and by the
Space Telescope Science Institute through grant AR-09926.01.
Development of the {\it STOKES} code was supported by the
Hans-B\"ockler-Stiftung.
\end{acknowledgements}

\bibliographystyle{aa}

\begin{thebibliography}{}

\bibitem[\protect\citeauthoryear{Balbus \&
Hawley}{1991}]{balbus_hawley91} Balbus, S. A., \& Hawley, J. F.
1991, ApJ, 376, 214

\bibitem[\protect\citeauthoryear{Baldwin et
al.}{1995}]{baldwin+95} Baldwin, J. A., Ferland, G. J., Korista, K.
T., \& Verner, D. A. 1995, \apj, 455, L119

\bibitem[\protect\citeauthoryear{Blumenthal \&
Mathews}{1975}]{blumenthal_mathews75} Blumenthal, G.~R., \& Mathews,
W.~G.\ 1975, \apj, 198, 517

\bibitem[\protect\citeauthoryear{Clavel et
al.}{1991}]{clavel+91} Clavel, J., et al.\ 1991, \apj, 366, 64

\bibitem[\protect\citeauthoryear{Crenshaw \& Blackwell}{1990}]{crenshaw_blackwell90}
Crenshaw, D.~M., \& Blackwell, J.~H., Jr.\ 1990, \apjl, 358, L37

\bibitem[Czerny et al.(2004)]{czerny+04} Czerny, B., Li, J.,
Loska, Z., \& Szczerba, R.\ 2004, \mnras, 348, L54

\bibitem[\protect\citeauthoryear{Dibai}{1981}]{dibai81} Dibai,
\'{E}.~A. 1981, Soviet Astron. Letts., 7, 248

\bibitem[\protect\citeauthoryear{Done \& Krolik}{1996}]{done_krolik96} Done, C., \& Krolik, J.~H.\ 1996,
\apj, 463, 144

\bibitem[\protect\citeauthoryear{Gaskell}{1982}]{gaskell82} Gaskell, C.~M.\ 1982,
ApJ, 263, 79

\bibitem[\protect\citeauthoryear{Gaskell}{1988}]{gaskell88} Gaskell,
C. M. 1988, ApJ, 325, 114

\bibitem[\protect\citeauthoryear{Gaskell}{2000}]{gaskell00} Gaskell, C.M.
2000, New Astron. Rev., 44, 564

\bibitem[\protect\citeauthoryear{Gaskell}{2008}]{gaskell08} Gaskell,
C. M. 2008, in The Nuclear Region, Host Galaxy, and Environment of
Active Galaxies, Rev. Mex. A\&A Conf. Ser. 32, 1

\bibitem[\protect\citeauthoryear{Gaskell \& Benker}{2008}]{gaskell_benker08} Gaskell, C.~M.~\& Benker, A.~J.~2008, ApJ,
~submitted [arXiv:0711.1013]

\bibitem[\protect\citeauthoryear{Gaskell \& Goosmann}{2008}]{gaskell_goosmann08} Gaskell, C. M. \& Goosmann, R.
W. 2008, ApJ submitted [arXiv:0805.4258]

\bibitem[\protect\citeauthoryear{Gaskell et al.}{2004}]{gaskell+04} Gaskell, C. M., Goosmann, R. W., Antonucci, R.,
\& Whysong, D. H. 2004, ApJ, 616, 147

\bibitem[\protect\citeauthoryear{Gaskell et al.}{2008a}]{gaskell+08a} Gaskell, C.~M., Goosmann, R.~W., Merkulova, N.~I., Shakhovskoy,
N.~M., \& Shoji, M.\ 2008a, ApJ Letters, submitted [arXiv:0711.1019]

\bibitem[\protect\citeauthoryear{Gaskell, Klimek, \&
Nazarova}{2008}]{gaskell+08b} Gaskell, C.~M., Klimek, E.~S., \&
Nazarova, L.~S.\ 2008b, ApJ, submitted [arXiv:0711.1025] (GKN)

\bibitem[\protect\citeauthoryear{Gaskell \& Sparke}{1986}]{gaskell_sparke86} Gaskell, C. M. \& Sparke, L. S.
1986, ApJ, 305, 175

\bibitem[\protect\citeauthoryear{Goosmann \& Gaskell}{2007}]{goosmann_gaskell07} Goosmann, R.~W., \& Gaskell, C.~M.\
2007, \aap, 465, 129

\bibitem[\protect\citeauthoryear{Hawley \& Krolik}{2001}]{hawley_krolik01} Hawley, J.~F., \&
Krolik, J.~H.\ 2001, \apj, 548, 348
%

\bibitem[\protect\citeauthoryear{Koratkar \& Blaes}{1999}]{koratkar_blaes99}
Koratkar, A. P. \& Blaes, O. 1999, \pasp, 111, 1

\bibitem[\protect\citeauthoryear{Koratkar \& Gaskell}{1989}]{koratkar_gaskell89} Koratkar, A. P. \& Gaskell, C.
M. 1989, ApJ, 345, 637

\bibitem[\protect\citeauthoryear{Koratkar \&
Gaskell}{1991a}]{koratkar_gaskell91a} Koratkar, A.~P., \& Gaskell,
C.~M.\ 1991a, \apj, 375, 85

\bibitem[\protect\citeauthoryear{Koratkar \&
Gaskell}{1991b}]{koratkar_gaskell91b} Koratkar, A.~P., \& Gaskell,
C.~M.\ 1991b, \apjs, 75, 719

\bibitem[\protect\citeauthoryear{Korista et al.}{1995}]{korista+95} Korista, K.\ T., et al.\
1995, \apjs, 97, 285

\bibitem[\protect\citeauthoryear{Leighly \&
Moore}{2004}]{leighly_moore04} Leighly, K.~M., \& Moore, J.~R.\
2004, \apj, 611, 107

\bibitem[MacAlpine(1972)]{macalpine72} MacAlpine, G.\ M.\ 1972,
\apj, 175, 11

\bibitem[\protect\citeauthoryear{MacAlpine}{1981}]{macalpine81} MacAlpine, G.~M.\ 1981,
\apj, 251, 465

\bibitem[\protect\citeauthoryear{MacAlpine}{2003}]{macalpine03} MacAlpine, G.~M.\
2003, RevMexAA (Ser. de Conf.), 18, 63

\bibitem[Netzer \& Laor(1993)]{netzer_laor93} Netzer, H., \& Laor,
A.\ 1993, \apjl, 404, L51

\bibitem[\protect\citeauthoryear{Osterbrock}{1978}]{osterbrock78} Osterbrock, D.~E.\ 1978, Proc. Natl. Acad. Sci., 75, 540

\bibitem[\protect\citeauthoryear{Peterson et al.}{1991}]{peterson+91}
Peterson, B. M. et al.\ 1991, \apj, 338, 119

\bibitem[\protect\citeauthoryear{Peterson et al.}{1992}]{peterson+92}
Peterson, B. M. et al.\ 1992, \apj, 392, 470

\bibitem[\protect\citeauthoryear{Smith et al.}{2004}]{smith+04} Smith, J.~E., Robinson,
A., Alexander, D.~M., Young, S., Axon, D.~J., \& Corbett, E.~A.\
2004, \mnras, 350, 140

\bibitem[Stoy(1933)]{stoy33} Stoy, R.\ H.\ 1933, \mnras, 93,
588

\bibitem[Suganuma et al.(2006)]{suganuma+06} Suganuma, M., et al.\
2006, \apj, 639, 46

\bibitem[\protect\citeauthoryear{Sulentic et al.}{2000}]{sulentic+00}
Sulentic, J.~W., Zwitter, T., Marziani, P., \& Dultzin-Hacyan, D.\
2000, \apjl, 536, L5

\bibitem[\protect\citeauthoryear{Welsh et al.}{2007}]{welsh+07} Welsh, W.~F., Martino,
D.~L., Kawaguchi, G., \& Kollatschny, W.\ 2007, in The Central
Engine of Active Galactic Nuclei, ed. L. C. Ho \& J.-M. Wang, ASP
Conf. Ser., 373, 29

\bibitem[\protect\citeauthoryear{Wilkes}{1984}]{wilkes84} Wilkes, B.~J.\ 1984, \mnras, 207, 73

\bibitem[\protect\citeauthoryear{Xu et al.}{2003}]{xu+03} Xu, D.~W., Komossa, S., Wei,
J.~Y., Qian, Y., \& Zheng, X.~Z.\ 2003, \apj, 590, 73

\end{thebibliography}

\end{document}